\documentclass[12pt]{article}
\usepackage{graphicx}
\usepackage{subfigure}

\def\napoli{University of G\"ottingen\\
G\"ottingen, GERMANY}
\def\support{\footnote{On behalf of the Belle Collaboration.}}

\def\Title#1{\begin{center} {\Large #1 } \end{center}}
\def\Author#1{\begin{center}{ \sc #1} \end{center}}
\def\Address#1{\begin{center}{ \it #1} \end{center}}

\newenvironment{Abstract}{\begin{quotation}  }{\end{quotation}}
\newenvironment{Presented}{\begin{quotation} \begin{center} 
             PRESENTED AT\end{center}\bigskip 
      \begin{center}\begin{large}}{\end{large}\end{center} \end{quotation}}

\def\beq{\begin{equation}}
\def\eeq#1{\label{#1}\end{equation}}
\def\eeqn{\end{equation}}

\def\beqa{\begin{eqnarray}}
\def\eeqa#1{\label{#1}\end{eqnarray}}
\def\eeqan{\end{eqnarray}}

\let\bar=\overbar

\def\Dslash{\not{\hbox{\kern-4pt $D$}}}
\def\dslash{\not{\hbox{\kern-2pt $\del$}}}

\def\msb{{\bar{\ssstyle M \kern -1pt S}}}

\begin{document}
\begin{titlepage}

\vfill
\Title{Charmless Semileptonic B Decays at $e^+e^-$ Colliders}
\vfill
\Author{ C\'esar Bele\~no\support}
\Address{\napoli}
\vfill
\begin{Abstract}
The following is an overview of the current status of charmless semileptonic $B$ decays and the extraction of the CKM matrix element $|V_{ub}|$ by the Belle and BaBaR collaborations. 
\end{Abstract}
\vfill
\begin{Presented}
Flavor Physics and CP Violation 2013\\
B\'uzioz, Rio de Janeiro, Brasil,  May 19--24, 2013
\end{Presented}
\vfill
\end{titlepage}
\def\thefootnote{\fnsymbol{footnote}}
\setcounter{footnote}{0}

\section{Introduction}

A precise measurement of the Cabibbo-Kobayashi-Maskawa (CKM) matrix element $|V_{ub}|$ is important to understand the nature of weak interactions and CP violation in the Standard Model~\cite{Petrella}. This value can be extracted from charmless semileptonic $B$ decays, where the final state hadron carries a $u$ quark and the lepton refers either to an electron or a muon. The advantage of these decays is that the leptonic and hadronic part of the final state do not interact strongly, and are thus easier to calculate theoretically. 

This presentation will summarize the status of measurements of charmless semileptonic decays and $V_{ub}$ from Belle and BaBaR experiments.

\section{Reconstruction of $B$ Mesons}

The study of $B$ mesons at $B$-factories are possible thanks to the tunning of the beam energies to the energy of the $\Upsilon(4S)$ resonance. The decay products of this resonance are mostly $B\bar{B}$ pairs, approximately 96\% of the cases, which allows to perform  precision measurements of $B$ decays. For this purpose, one $B$ meson is reconstructed in the decay mode of interest, the signal, by combining a lepton, a charmless meson and a neutrino. Since the latter is not visible to the detector, it is inferred from the missing four-momentum, 
\begin{equation}
 P_{\rm miss} = P_{\rm beam}-\sum_{i}P_{i},
\end{equation}
which is basically the difference between the beam four-momentum and the sum of all four-momenta of the reconstructed particles. 

Nowadays, there are three methods for reconstructing $B$ mesons~\cite{Dingfelder}, namely untagged, semileptonic tag and hadronic tag. In the untagged method, one only reconstruct the signal $B$, which offers a big statistical sample of signal candidates, but also incurs in a lot much bigger amount of background. Therefore, the signal reconstruction efficiency is high but the signal purity is poor. One of the challenges is the reduction of the background due to charmed semileptonic decays that is about 50 times more abundant than charmless semileptonic decays and the kinematic of the processes are very similar. Consequently, one has to apply very harsh selection on kinematic variables to suppress this background. The selection of a $B$ meson candidate is evaluated with two variables, the beam constrained mass $M_{bc}$ and the energy difference $\Delta E$ given by,
\begin{equation}
 M_{\rm bc} = \sqrt{E_{\rm beam}^{* 2 } - |\vec{p}_{B}^{*}|^{2}}, 
\end{equation}
\begin{equation}
 \Delta E = E_{\rm beam}^{*}- E_{B}^{*2}, 
\end{equation}
where $E_{\rm beam}^{*}$ is the energy of the beam and $P_{B}^{*}$ and $E_{\rm B}^{*}$ are the momentum and energy of the $B$ meson in the $\Upsilon(4S)$ rest frame. 

The other two techniques use information about the other $B$ meson or $B_{\rm tag}$, this has a direct effect on the signal sample, the signal purity is increased. However, there is a price to pay, the reconstruction efficiency is dramatically lowered. In the semileptonic tag technique~\cite{Hokuue}, the $B_{\rm tag}$ is reconstructed in charmed semileptonic decays of the form $B\to D^{(*)}\ell\nu$, with the $D^{(*)}$ meson reconstructed in hadronic modes. Since there are two semileptonic decays involved, additional requirements have to be considered. Foremostly, one identifies two leptons of opposite signs and then consider the presence of a neutrino on each side, following the massless neutrino hypothesis in the $\Upsilon(4S)$ rest frame, i.e., $P_{\nu}^{*2}=0=(P_{B}^{*}-P_{Y}^{*})^2$, where $P_{B}^{*}$ and $P_{Y}^{*}$ are the four-momentum of the $B$ meson and a virtual particle $Y$ respectively and all asterisked quantities are boosted to the $\Upsilon(4S)$ resonance.  The $Y$ particle is formed by the combination of a charged lepton and the reconstructed meson. From this condition, one can infere the angle between the $B$ meson and the $Y$ system through $\cos \theta_{BY} = \frac{2E_{B}^{*}E_{Y}^{*}-m_{B}^{2}m_{Y}^{2}}{2|\vec{p}_{B}^{*}||\vec{p}_{Y}^{*}|}$. This quantity can be used for the tag side, $\cos\theta_{B1}
$, the signal side, $\cos\theta_{B2}$, together with the cosine of the angle between the two $Y$ particles on each side, $\cos\theta_{12}$, to form a discriminating variable to separate signal events from background events $x_{B}^{2}$ ( as is called within the Belle Collaboration or $\cos^2\phi_{B}$ for the case of BaBaR) given by
\begin{equation}
x_{B}^{2}=1-\frac{1}{\sin^2 \theta_{12}}\left( \cos^2\theta_{B1} + \cos^2\theta_{B2}-2\cos\theta_{B1}cos\theta_{B2}\cos\theta_{12}\right), 
\end{equation}
that for signal events this quantity has values in the interval $[0,1]$.

Finally, in the hadronic tag method the $B_{\rm tag}$  is fully reconstructed in hadronic decays leading to the knowledge of the momentum, flavor and charge of the signal $B$ and also offers a very high signal purity. Discriminating variables for this technique include the missing mass squared $M_{\rm miss}^2$ that is basically the squared magnitud of the missing four-momentum, $M_{\rm miss}^{2}=|P_{\rm miss}|^{2}$, the hadronic invariant mass $m_{X}$ and the squared of the momentum transfer to the lepton $q^{2}=(P_{\ell}+P_{\nu})^2$. The significant high purity achieved with this technique leads to a high reduction of systematic uncertainties, but it also requires a large dataset to reduce statistical uncertainties.

\section{Inclusive Charmless Semileptonic $B$ Decays}

In the inclusive approach the meson $X_{u}$ in the semileptonic decay $B\to X_{u}\ell\nu$ is modeled as a sum over all hadronic final states that carry an $u$ quark~\cite{Jin}. At parton level, the decay rate can be computed from a free quark decay as $\Gamma(b\to u\ell\nu)=\frac{ G_{F}^{2}m_{b}^{5} }{ 192\pi^{3}} |V_{ub}|^{2} $, where $G_{F}$ is the Fermi constant, $m_{b}$ the mass of the $b$ quark and $V_{ub}$ is the CKM matrix element for $b\to u$ quark transition. However, the $b$ quark is not a free particle, instead it forms bound states inside the $B$ mesons, for which other considerations need to be taken into account, such as strong interaction effects or the movement of $b$ quarks inside B mesons. The Operator Product Expansion (OPE) and the Heavy Quark Expansion (HQE) are used to calculate the total decay rate for $B\to X_{u}\ell\nu$ in powers of $1/m_{b}$ with uncertainties at the $5\%$ level. The additional factors to the free quark decay correspond to electroweak, perturbative and non-perturbative QCD corrections. Hence the decay rate can be written as~\cite{AndersenGardi}:
\begin{equation}
\Gamma(B\to X_{u}\ell\nu) = \frac{ G_{F}^{2}m_{b}^{5} }{ 192\pi^{3} } |V_{ub}|^{2} \left[ 1 + a_{0}\frac{ \alpha_{s}(m_{b}) }{ \pi }+a_{1}\left( \frac{ \alpha_{s}(m_{b}) }{ \pi }\right)^2 +\dots + {\cal O}\left(\frac{ \Lambda^{2} }{ m_{b}^{2}}\right) \right],
\end{equation}
where $\alpha_{s}$ and $\Lambda$ are the strong and QCD coupling constants. 

The application of very harsh selection on kinematic variables for reducing the large background from $B\to X_{c}\ell\nu$, reduce the available phase space region. This represents a challenge on the theory side since the HQE convergence is spoiled and non-perturbative Shape Functions (SFs) need to be introduced~\cite{Luth}. The configuration of the SFs cannot be determined from first principles, instead they rely on global fits to moments in inclusive $B\to X_{c}\ell\nu$ and $B\to X_{s}\gamma$. From these fits heavy quark parameters can be determined, such as the mass of the $b$ quark $m_{b}$, the kinetic energy squared of the $b$ quark in the $B$ meson $\mu_{b}^{2}$ and the chromomagnetic moment $\mu_{G}^{2}$. Due to confinement and non-perturbative effects, the quantitative values of these parameters rely upon the theoretical framework in which they are defined. Therefore the results of the global fits need to be translated to other schemes depending on the QCD calculation used to extract $|V_{ub}|$.

Some of the QCD calculations available to date are based on either OPE such as BLNP~\cite{BLNP} and GGOU~\cite{GGOU}, or non-perturbative QCD models like Dressed Gluon Exponentiation (DGE)~\cite{AndersenGardi} and ADFR~\cite{ADFR}. The main uncertainty in these calculations is coming from the uncertainty in the $m_{b}$. In the case of the ADFR calculation the dominant uncertainty is due to the uncertainty in the mass of the charm quark $m_{c}$.

 \begin{figure}[htb]
 \centering
 \includegraphics[height=1.5in]{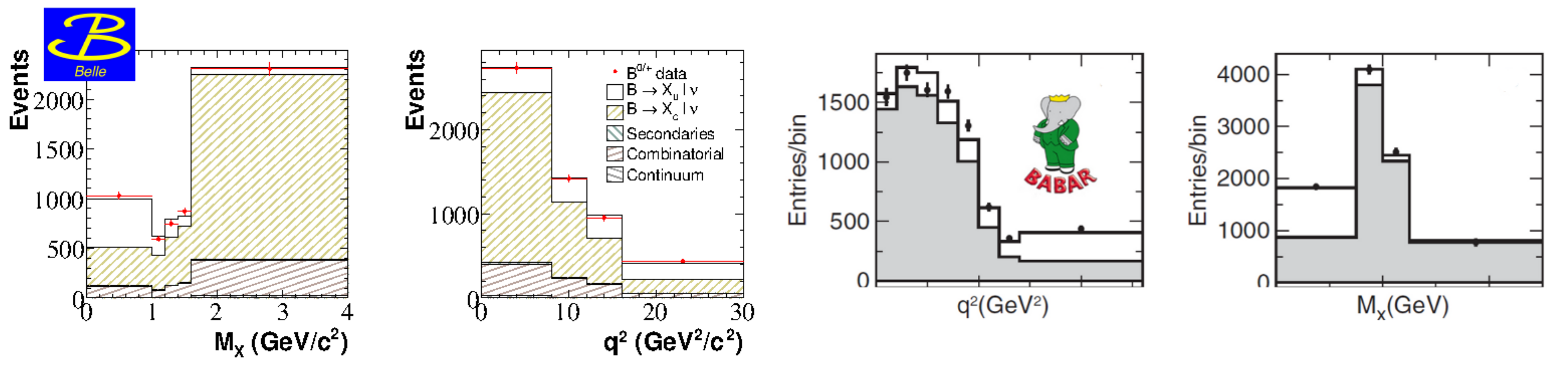}
 \caption{Fit Projections for $M_{X}$ and $q^{2}$. Left: results from Belle. Right: results from BaBaR.}
 \label{fig:inclusive}
 \end{figure}

The most recent results on inclusive measurements from Belle~\cite{InclusiveBelle} and BaBaR~\cite{InclusiveBabar} use their complete data sample, $467\times 10^{6}$ $B\bar{B}$ pairs for BaBaR and $657\times 10^{6}$ $B\bar{B}$ pairs for Belle, and a hadronic tag technique to reconstruct the $B_{\rm tag}$. The two analyses differ in the treatment of the background, Belle~\cite{InclusiveBelle} implements a multivariate discriminant while BaBaR~\cite{InclusiveBabar} uses a cut based method for background suppression. BaBaR also reports measurements of $|V_{ub}|$ in several regions of phase space. The most precise results are extracted from a two dimensional fit to $(M_{X},q^{2})$ with no restriction on phase space other than $p_{\ell}^{*}>1.0$ GeV, this selection allows to access approximately $90\%$ of the total phase space. The distributions of the projections of the fit for $M_{X}$ and $q^{2}$ are shown in Fig.~\ref{fig:inclusive}. The upper row corresponds to the projections as measured by the Belle Collaboration, for which $1032\pm 91$ signal events are extracted leading to a partial branching fraction of $\Delta {\cal B} (B\to X_{u}\ell\nu)=(1.96 \pm 0.17_{\rm stat} \pm 0.16_{\rm syst})\times 10^{-3}$. The second row shows the projections from the BaBaR collaboration with $1430\pm 130$ signal events and the corresponding partial branching ratio of $\Delta {\cal B} (B\to X_{u}\ell\nu)=(1.80 \pm 0.13_{\rm stat} \pm 0.15_{\rm syst})\times 10^{-3}$.

The values of $|V_{ub}|$ can be extracted using the relation~\cite{InclusiveBabar}
\begin{equation}
\label{vubeq}  
|V_{ub}|=\sqrt{ \frac{ \Delta {\cal B} (B\to X_{u}\ell\nu )}{ \tau_{B} \Gamma_{\rm theory}} },
\end{equation}
where $\tau_{B}$ is the $B$ lifetime and $\Delta {\cal B} (B\to X_{u}\ell\nu )$ and $\Gamma_{\rm theory}$  are the measured partial branching ratio and the predicted decay rate in a given phase space region, respectively. The latter depends on the QCD calculation implemented. The inclusive $|V_{ub}|$ extracted with four different QCD calculations for the Belle and BaBaR collaborations together with the world average from HFAG  are shown in Table~\ref{tab:inclusive}.  

\begin{table}[t]
\begin{center}
\begin{tabular}{l|cccc}  \hline
             &  BLNP~\cite{BLNP}  &  DGE~\cite{AndersenGardi} & GGOU~\cite{GGOU} & ADFR~\cite{ADFR} \\ \hline
Belle~\cite{InclusiveBelle}        &$4.47\pm 0.27^{+0.19}_{-0.21}$&$4.60\pm 0.27^{+0.11}_{-0.13}$&$4.54\pm 0.27^{+0.10}_{-0.11}$&$4.48\pm 0.30^{+0.19}_{-0.19}$  \\
BaBaR~\cite{InclusiveBabar}        &$4.28\pm 0.24^{+0.18}_{-0.20}$&$4.40\pm 0.24^{+0.12}_{-0.13}$&$4.35\pm 0.24^{+0.09}_{-0.10}$&$4.29\pm 0.24^{+0.18}_{-0.19}$  \\ 
World average~\cite{HFAG}&$4.40\pm 0.15^{+0.19}_{-0.21}$&$4.45\pm 0.15^{+0.15}_{-0.16}$&$4.39\pm 0.15^{+0.12}_{-0.14}$&$4.03\pm 0.13^{+0.18}_{-0.12}$  \\ \hline
\end{tabular}
\caption{Values of $|V_{ub}|$ reported by the Belle and BaBaR collaboration compared to the world average, where the first error is statistical and the second due to systematic uncertainties. The values are extracted using four different QCD calculation with the requirement $p_{\ell}^{*}>1.0$ GeV.}
\label{tab:inclusive}
\end{center}
\end{table}

\section{Exclusive Charmless Semileptonic $B$ Decays}
In the exclusive approach the hadronic final state is reconstructed in a particular channel such as $B\to\pi\ell\nu$ or $B\to\rho\ell\nu$. In this case the matrix element necessary for calculating the decay rate depends not only on the CKM element $V_{ub}$ but also on non-perturbative hadronic physics contained in so called form factors. These form factors depend on the final state hadron, in particular on whether the particle is a vector or a pseudoscalar meson. For vector mesons, such as $\rho$ and $\omega$, the decay rate can be written as~\cite{BabarUntaggedW}
\begin{equation}
\frac{ d\Gamma(B\to V\ell\nu)  }{ dq^{2} } = \frac{ G_{F}^{2} p_{V} q^{2} }{ 96\pi^{3}m_{B} C_{V}^{2} } |V_{ub}|^{2} (|H_{0}|^2+|H_{+}|^2+|H_{-}|^2)
\end{equation}
where $G_{F}$ is the Fermi constant, $p_{V}$ is the magnitude of the vector meson momentum in the $B $ rest frame, $m_{B}$ is the $B$ mass, $C_{V}$ is the isospin factor and $H_{\pm}$ and $H_{0}$ are the helicity functions. These helicity functions can be written in terms of two axial vector form factors $A_{1}$ and $A_{2}$, and one vector form factor $V$ as follows
\begin{equation}
H_{\pm}(q^2)=(m_{B}+m_{V})\left[ A_{1}(q^2) \mp \frac{ 2m_{B}p_{V} }{ (m_{B}+m_{V})^2 }V(q^{2}) \right],
\end{equation}
\begin{equation}
H_{0}(q^2)= \frac{ m_{B}+m_{V} }{ 2m_{V}\sqrt{q^{2}} }\left[ (m_{B}^{2}-m_{V}^{2}-q^{2})A_{1}(q^{2}) -\frac{ 4m_{B}^{2}p_{V}^{2} }{ (m_{B} +m_{V})^{2} }A_{2}(q^{2}) \right].
\end{equation}
For pseudoscalar mesons the decay rate only depends on one form factor, because in the limit of small lepton masses the term proportional to the second form factor $f_{0}(q^{2})$ can be neglected, hence~\cite{Luth,BelleHad}
\begin{equation}
\frac{ d\Gamma(B\to P\ell\nu)  }{ dq^{2} } = \frac{ G_{F}^{2} |p_{P}|^{3}  }{ 24\pi^{3} } |V_{ub}|^{2} |f_{+}(q^{2})|^{2}, 
\end{equation}
with $f_{+}(q^{2})$ a form factor. These form factors are provided by theory using different calculations valid for certain regions of $q^{2}$. These calculations include quark models like ISGW2, Lattice QCD (LQCD) that is valid for high values of $q^{2}$ and Light Cone Sum Rules (LCSR) valid for low values of $q^{2}$. 

\subsection{Results for $B\to\pi\ell\nu$}
Among the exclusive channels studied to date, the $B\to\pi\ell\nu$ is the most promising channel to extract the CKM matrix element $|V_{ub}|$, both theorically and experimentally. Different analyses have come out using different reconstruction techniques of $B$ mesons, taking advantage of the large signal sample, with consistent results on the values on $|V_{ub}|$.
\begin{figure}[htb]
\centering
\subfigure[Belle]{
\includegraphics[width=0.45\textwidth ]{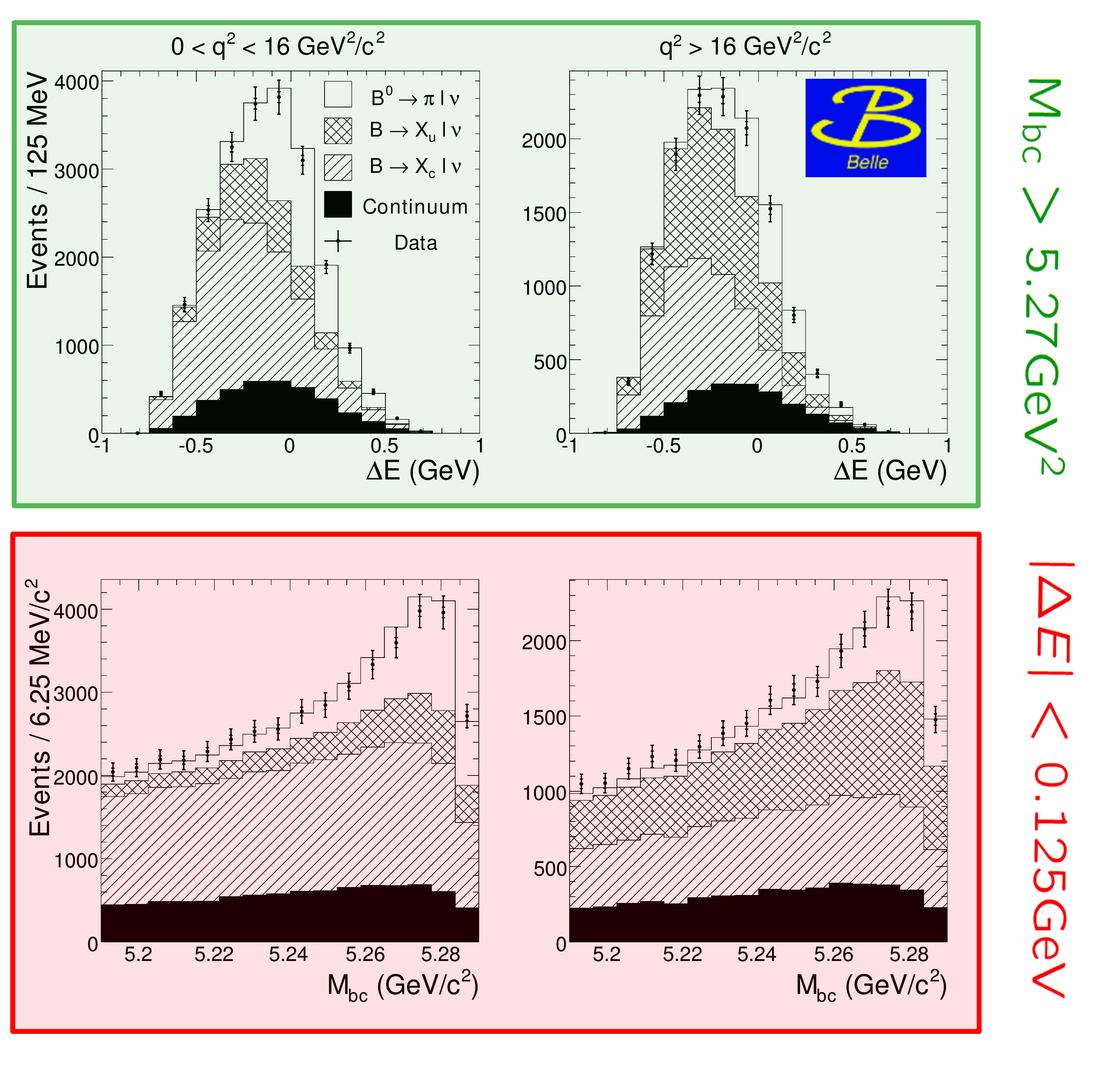}
\label{untaggedbelle}
}
\subfigure[BaBaR]{
\includegraphics[width=0.45\textwidth ]{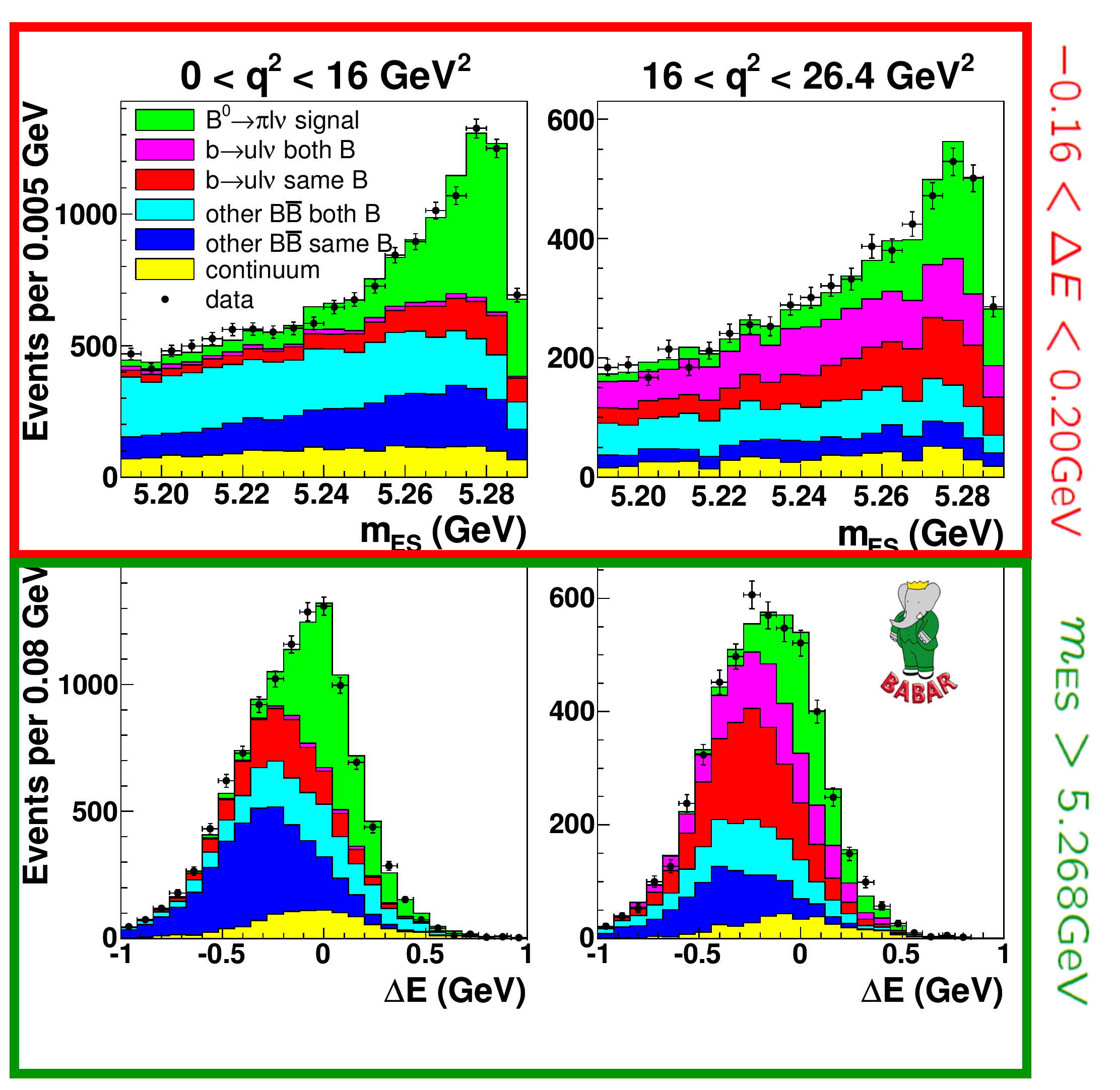}
\label{untaggedbabar}
}
\caption{Fit projections in two bins of $q^{2}$ for the $\Delta E$ and $M_{\rm bc(ES)}$ from the untagged measurements of $B\to\pi\ell\nu$ by the Belle and BaBaR collaborations.}
\end{figure}
The most recent untagged results from Belle~\cite{UntaggedExclusiveBelle} and BaBaR~\cite{UntaggedExclusiveBabar} perform a two dimensional binned extended maximum likelihood fit to $(M_{bc(ES)},\Delta E)$, in bins of $q^{2}$ and consider basically three sources of background: charmless semileptonic $B$ decays other than the signal $(B\to X_{u}\ell\nu)$, other $B$ decays specially $B\to X_{c}\ell\nu$ and continuum events.  BaBaR splits further the $B$ decays into  `same $B$ category' and `both $B$ category' depending wether or not the pion and the lepton come from the same $B$ meson. In both analyses, the selection criteria are optimized separately in each bin of $q^{2}$ by maximizing the figure of merit $S/\sqrt{S+B}$, where $S(B)$ is the expected number of signal (background) events. In addition, an unfolding technique is performed to correct for finite resolution and acceptance, by applying the inverse detector response matrix to the measured yields.  The Belle~\cite{UntaggedExclusiveBelle} analysis uses a data sample corresponding to $605$ fb$^{-1}$  of integrated luminosity, for which $21486\pm 548$ signal events are obtained, leading to a branching ratio of ${\cal B}(B^{0}\to \pi^{-}\ell^{+}\nu) = (1.49\pm 0.04_{\rm stat}\pm 0.07_{\rm syst} )\times 10^{-4}$.   The BaBaR~\cite{UntaggedExclusiveBabar} analysis reconstruct explicitly the $B^{0}\to \pi^{-}\ell^{+}\nu$ and $B^{+}\to \pi^{0}\ell^{+}\nu$ decay channels and combine the results using isospin symmetry, for which they report $12448\pm 361$ signal events leading to a branching ratio of ${\cal B} (B\to\pi\ell\nu)=(1.45\pm 0.04_{\rm stat}\pm 0.06_{\rm syst})\times 10^{-4}$. The projections of the fit result in $\Delta E$ and $M_{\rm bc(ES)}$ are shown in Fig.~\ref{untaggedbelle} for the Belle analysis and in Fig.~\ref{untaggedbabar} for the BaBaR analyisis. The main sources of systematic uncertainties are due to detector effects. 

\begin{figure}[htb]
\centering
\subfigure[BaBaR]{
\includegraphics[width=0.25\textwidth ]{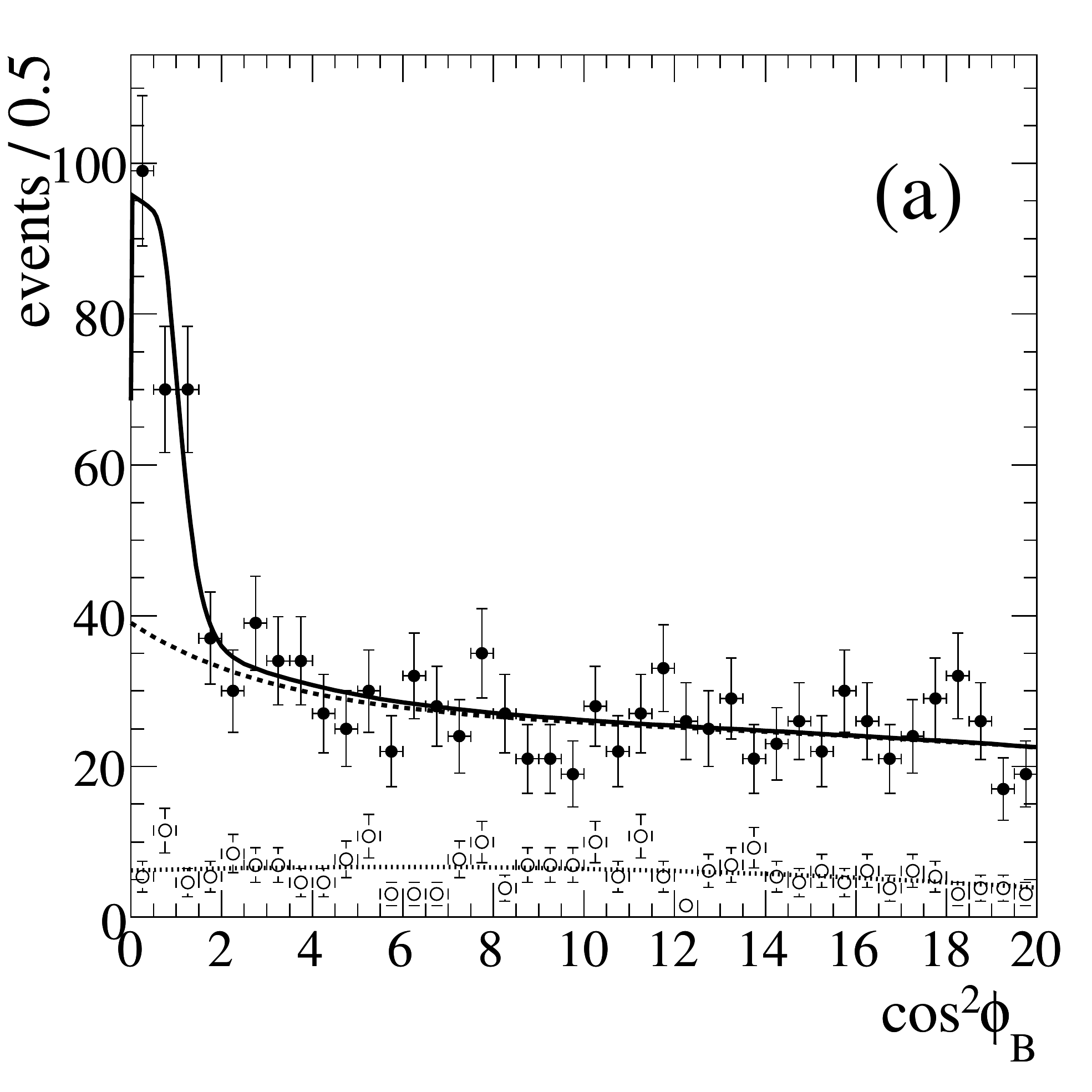}
}
\subfigure[Belle]{
\includegraphics[width=0.45\textwidth ]{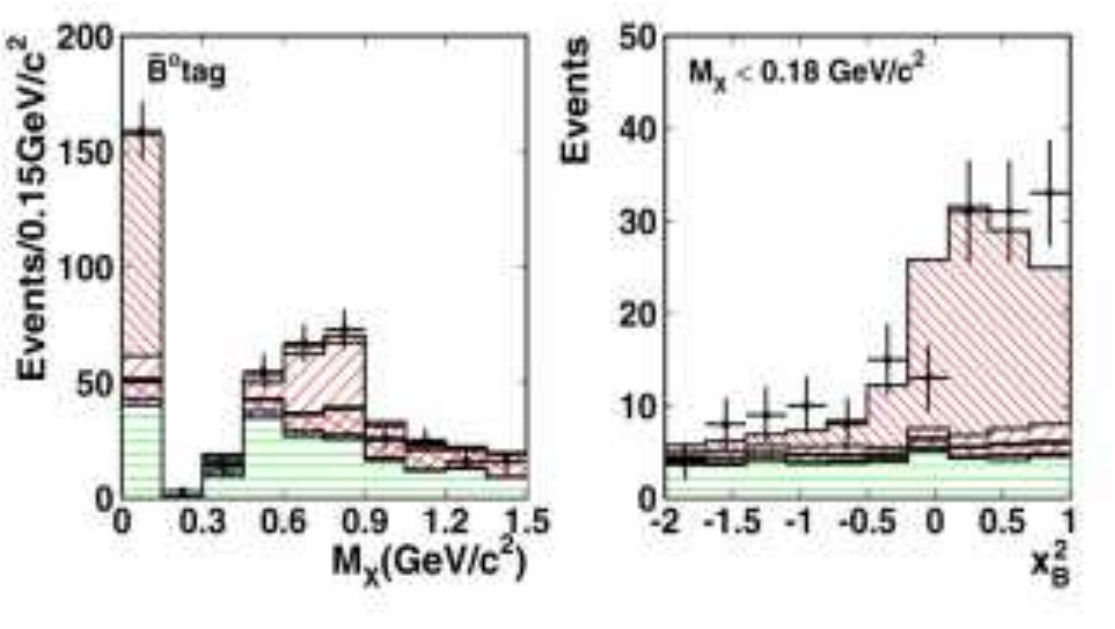}
}
\label{fig:SLtag}
\caption{Fit results from the semileptonic tag measurements of $B\to\pi\ell\nu$ by the Belle and BaBaR collaborations.}
\end{figure}

The lattest results from Belle and BaBaR using semileptonic tag reconstruction differ in the method use to extract the signal yields, while BaBaR~\cite{SLBabar} uses an unbinned maximum likelihood to $cos^{2}\phi_{B}$ distribution, Belle~\cite{Hokuue} implements a two dimensional binned maximum likelihood to the $(x_{B},m_{X})$ distributions. The fit results for $B^{0}\to\pi^{-}\ell^{+}\nu$ are shown in Fig.~\ref{fig:SLtag} and the measurements are unfolded. The analysis by Belle reports $156\pm 20$ $B^{0}\to\pi^{-}\ell^{+}\nu $ events using a data set of $253$ fb$^{-1}$, which leads to a branching ratio of  ${\cal B}(B^{0}\to \pi^{-}\ell^{+}\nu) = (1.38\pm 0.19_{\rm stat}\pm 0.14_{\rm syst} )\times 10^{-4}$. For the same decay channel, BaBaR reports $150\pm 22$ signal events leading to a branching ratio of  ${\cal B}(B^{0}\to \pi^{-}\ell^{+}\nu) = (1.38\pm 0.21_{\rm stat}\pm 0.07_{\rm syst} )\times 10^{-4}$ using a data sample of 348 fb$^{-1}$ of integrated luminosity.    

\begin{figure}[htb]
\centering
\subfigure[$B^{0}\to\pi^{-}\ell^{+}\nu$]{
\includegraphics[width=0.45\textwidth ]{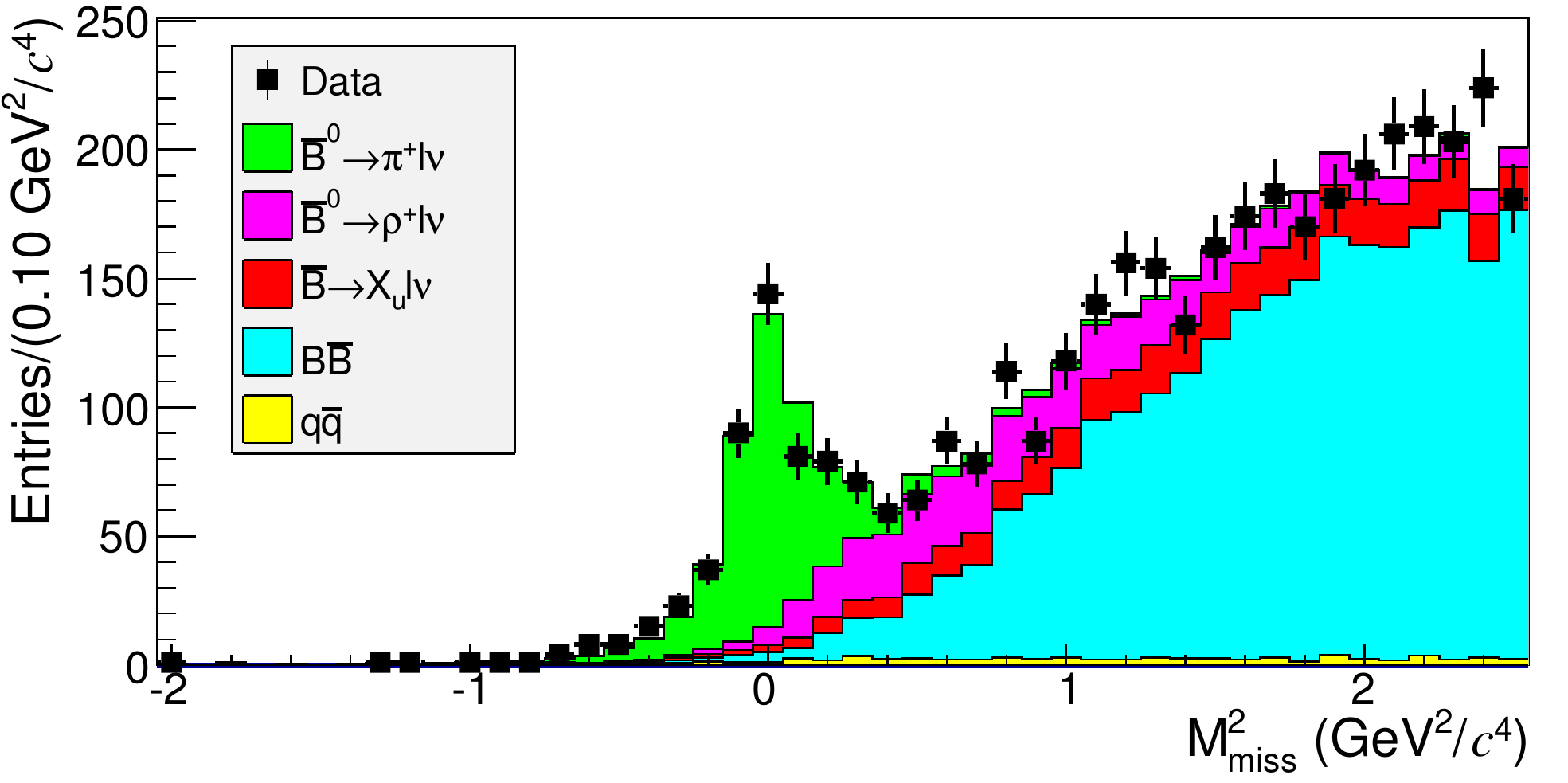}
}
\subfigure[$B^{+}\to\pi^{0}\ell^{+}\nu$]{
\includegraphics[width=0.45\textwidth ]{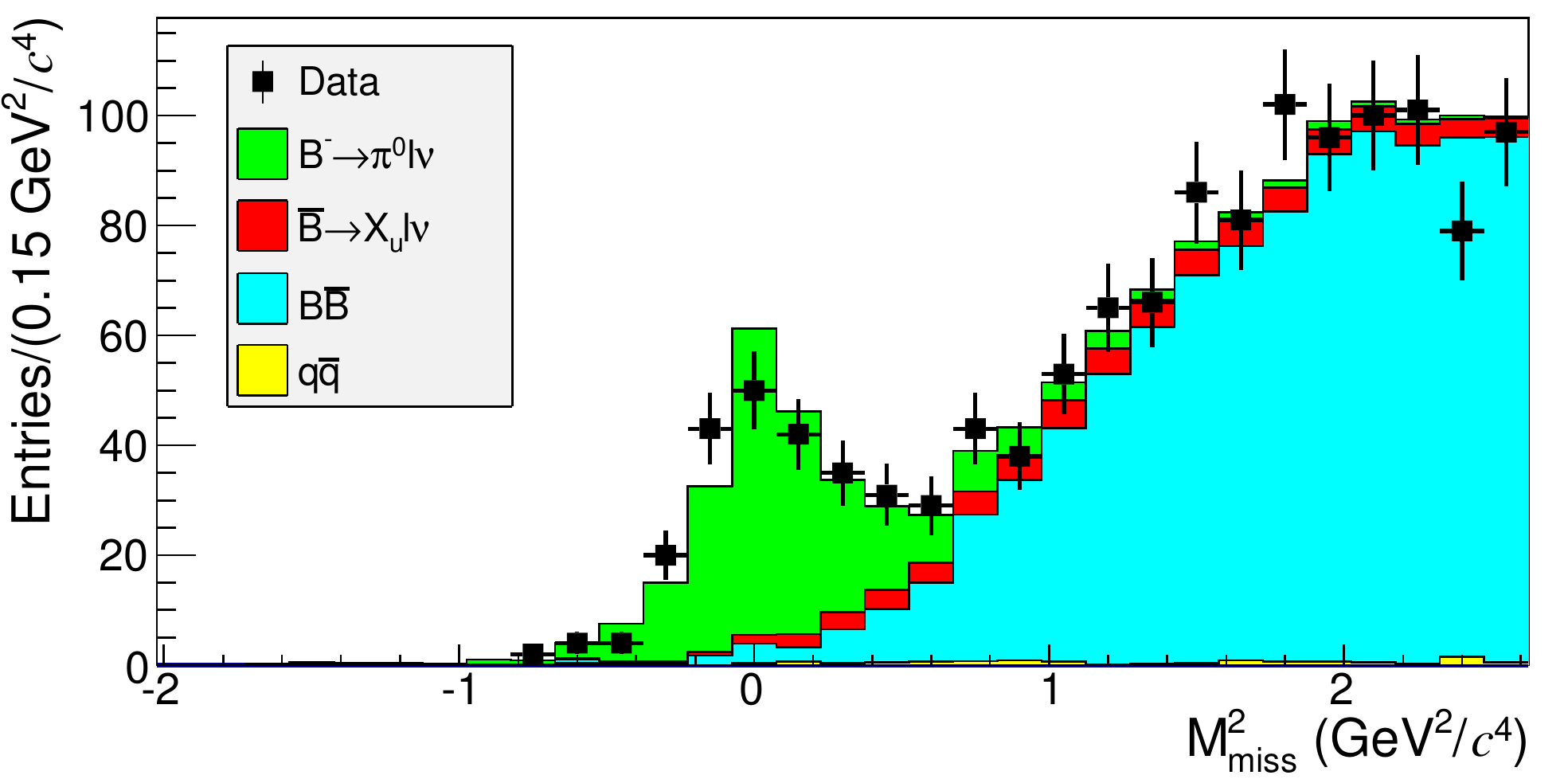}
}
\label{fig:hadBelle}
\caption{Fit results from the hadronic tag measurements of $B\to\pi\ell\nu$ from Belle.}
\end{figure}

A new analysis using hadronic reconstruction by Belle~\cite{BelleHad} obtains $463\pm 28$ $B^{0}\to\pi^{-}\ell^{+}\nu$ events from a extended binned maximum likelihood fit to the $M_{\rm miss}^{2}$ distribution, using a data set of 711 fb$^{-1}$. This leads to a branching ratio of $ {\cal B}(B^{0}\to \pi^{-}\ell^{+}\nu) = (1.49\pm 0.09_{\rm stat}\pm 0.07_{\rm syst} )\times 10^{-4}$, which is competive with the more precise results from untagged measurements. Fig.~\ref{fig:hadBelle} shows the fitted $M_{\rm miss}^{2}$ distribution for $B^{0}\to\pi^{-}\ell\nu$ and $B^{+}\to\pi^{0}\ell\nu$, it can be noted a clear peak around $M_{\rm miss}^2 = 0 $ GeV$^{2}$ with a small contribution from the background underneath this peak. The branching ratio for $B^{+}\to\pi^{0}\ell^{+}\nu$ channel is measured to be $ {\cal B}(B^{+}\to \pi^{0}\ell^{+}\nu) = (0.80\pm 0.08_{\rm stat}\pm 0.04_{\rm syst} )\times 10^{-4}$, which is in good agreement with the predictions using isospin symmetry, $2\times \frac{ {\cal B}(B^{+}\to\pi^{0}\ell^{+}\nu) }{ {\cal B}(B^{0}\to\pi^{-}\ell^{+}\nu) } \frac{ \tau_{ B^{0} } }{ \tau_{ B^{+} } }=1.00\pm 0.13_{\rm tot}$, using $ \frac{ \tau_{ B^{+} } }{ \tau_{ B^{} } }=1.079\pm 0.007$~\cite{pdg} .  The reduced contribution of background events in the hadronic tag technique is a direct consequence of the kinematic reconstruction of the full event and also contributes to the reduction of systematic uncertainties.

To extract a value for $|V_{ub}|$ from the measured differential decay rates, the traditional method described by Eq.~\ref{vubeq} is used with a slight modification. $\Delta {\cal B}$ is replaced by $C_{V}\Delta {\cal B}$, where the isospin factor $C_{V}$ is equal to 2 for $B^{+}$ decay modes and is equal to 1 for $B^{0}$ decays modes. A total of four form factor calculations are used to determined $V_{ub}$, two from unquenchend LQCD valid for $q^{2}>$ 16 GeV$^{2}$ provided by the FNAL~\cite{FNAL} and HPQCD~\cite{HPQCD} collaborations, and two from LCSR provided by Ball and Zwicky~\cite{Ball} ($q^{2}<16$ GeV$^{2}$) and KMOW~\cite{KMOW} ($q^{2}<12$ GeV$^2{2}$). I only present the most recent $V_{ub}$ results from the untagged BaBaR and the tagged Belle analysis and compare with the world averages according to HFAG~\cite{HFAG} (end of 2011), which are shown in Table~\ref{tab:exclusive}.   

\begin{table}[t]
\begin{center}
\begin{tabular}{l|ccc}  \hline
             &  Belle tagged~\cite{BelleHad}  &  BaBaR Untagged~\cite{UntaggedExclusiveBabar} & World Average~\cite{HFAG} \\ \hline
KMOW~\cite{KMOW}   &$3.40\pm 0.13\pm0.09^{+0.37}_{-0.32}$&$3.46\pm 0.06\pm 0.08^{+0.37}_{-0.32}$&$3.40\pm 0.07^{+0.37}_{-0.32}$  \\
Ball-Zwicky~\cite{Ball} &$3.58\pm 0.12\pm 0.09^{+0.59}_{-0.39}$&---&$3.57\pm 0.06^{+0.59}_{-0.39}$  \\ 
HPQCD~\cite{HPQCD} &$3.81\pm 0.22\pm 0.10^{+0.66}_{-0.43}$&$3.47\pm 0.10\pm 0.08^{+0.60}_{-0.39}$&$3.45\pm 0.09^{+0.60}_{-0.39}$  \\ 
FNAL~\cite{FNAL}&$3.64\pm 0.21\pm 0.09^{+0.40}_{-0.33}$&$3.31\pm 0.09\pm 0.07^{+0.37}_{-0.30}$&$3.30\pm 0.09^{+0.37}_{-0.30}$  \\ \hline
\end{tabular}
\caption{Values of $|V_{ub}|$ from $B^{0}\to\pi^{-}\ell^{+}\nu$ reported by the Belle and BaBaR collaboration compared to the world average, where the first error is statistical, the second due to systematic uncertainties and the third is the theoretical uncertainty. }
\label{tab:exclusive}
\end{center}
\end{table}

\begin{figure}[htb]
\centering
\subfigure[BaBaR]{
\includegraphics[width=0.35\textwidth ]{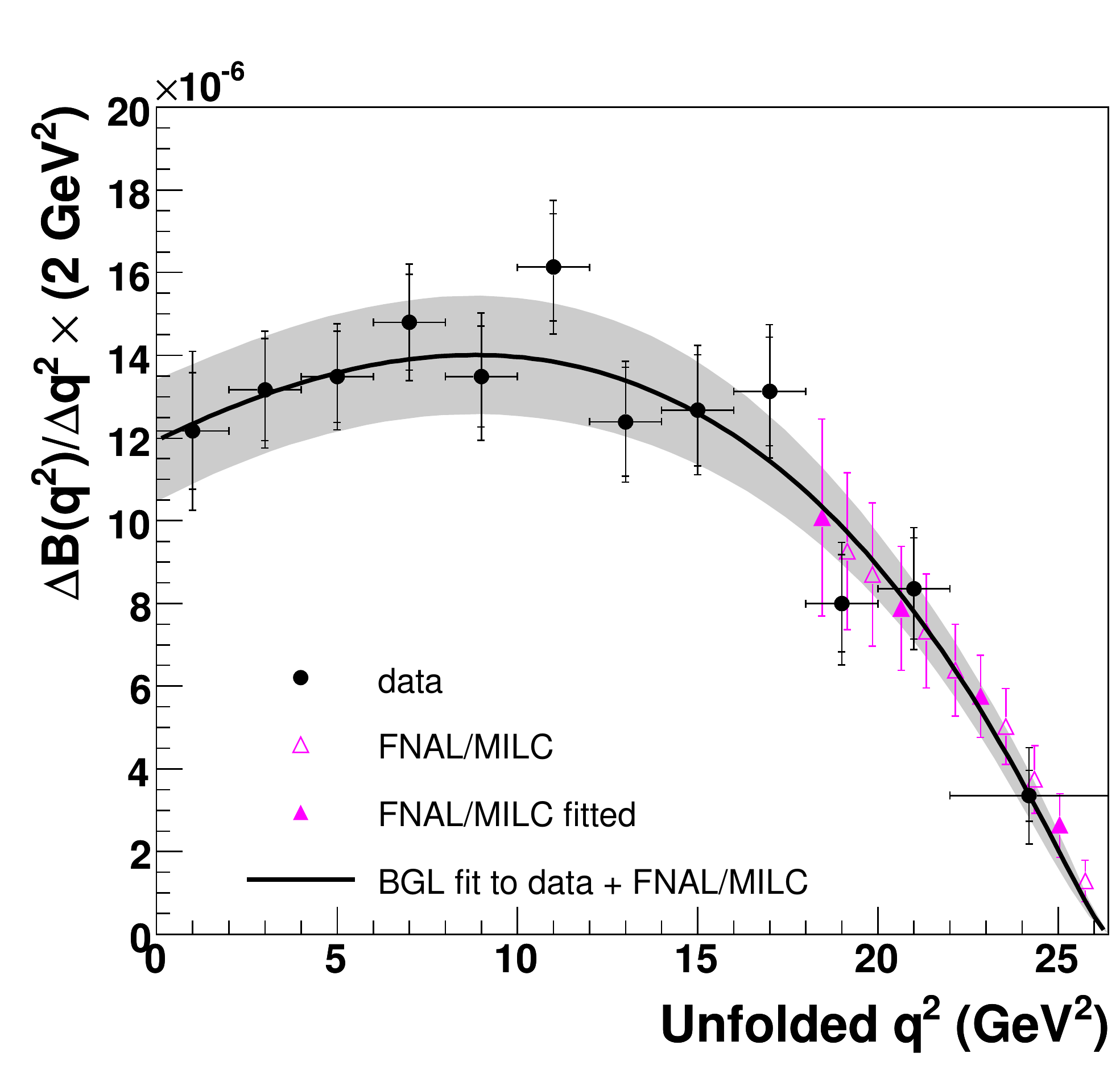}
\label{fig:q2BRbabar}
}
\subfigure[Belle]{
\includegraphics[width=0.55\textwidth ]{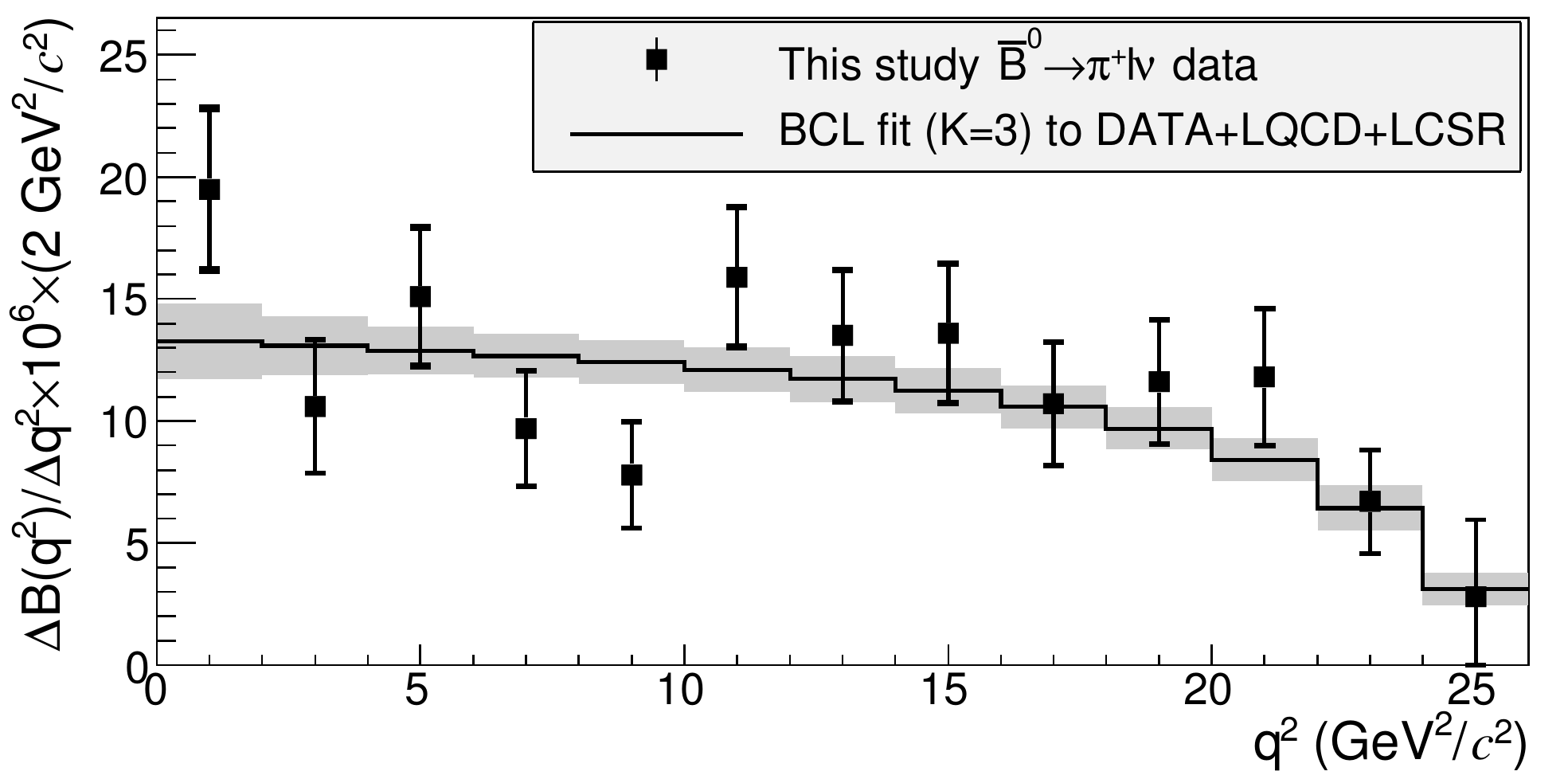}
\label{fig:q2BRbelle}
}
\caption{Simultaneous fit of form factor parametrization to $B^0\to\pi^-\ell^+\nu$ data, (a) using the BGL parametrization~\cite{BGL} to the BaBaR data~\cite{UntaggedExclusiveBabar} and four points of the FNAL~\cite{FNAL} prediction and (b) using the BCL parametrization~\cite{BCL} to the Belle data~\cite{BelleHad} using LQCD points and LCSR prediction at $q^2 = 0$. }
\end{figure}

A second method to extract $V_{ub}$ consists in applying a simultaneous fit using a model independent description of the $f_{+}(q^{2})$ hadronic form factor and the measured $q^{2}$ spectrum. The BaBaR~\cite{UntaggedExclusiveBabar} collaboration performs a simultaneous fit of the BGL parametrization~\cite{BGL} to their experimental data and to four points of the  FNAL predictions for the $B^{0}\to\pi^{-}\ell\nu$ to obtain $|V_{ub}|=(3.25\pm 0.31)\times 10^{-3}$ (see Fig.~\ref{fig:q2BRbabar}). The Belle~\cite{BelleHad} collaboration reports a value of $|V_{ub}|=(3.52\pm 0.29)\times 10^{-3}$ using the BCL parametrization~\cite{BCL} with their measured partial branching fractions, a recent LCSR calculation and LQCD points (see Fig.~\ref{fig:q2BRbelle}). Belle also applies the same procedure using data from previous untagged measurements by Belle~\cite{UntaggedExclusiveBelle} and BaBaR~\cite{UntaggedExclusiveBabar} with their tagged~\cite{BelleHad} results and obtain a value of $|V_{ub}|=(3.41\pm 0.22)\times 10^{-3}$. 

\subsection{Results from other exclusive channels}
\subsubsection{$B\to\rho\ell\nu$}
 \begin{figure}[htb]
 \centering
 \includegraphics[height=2in]{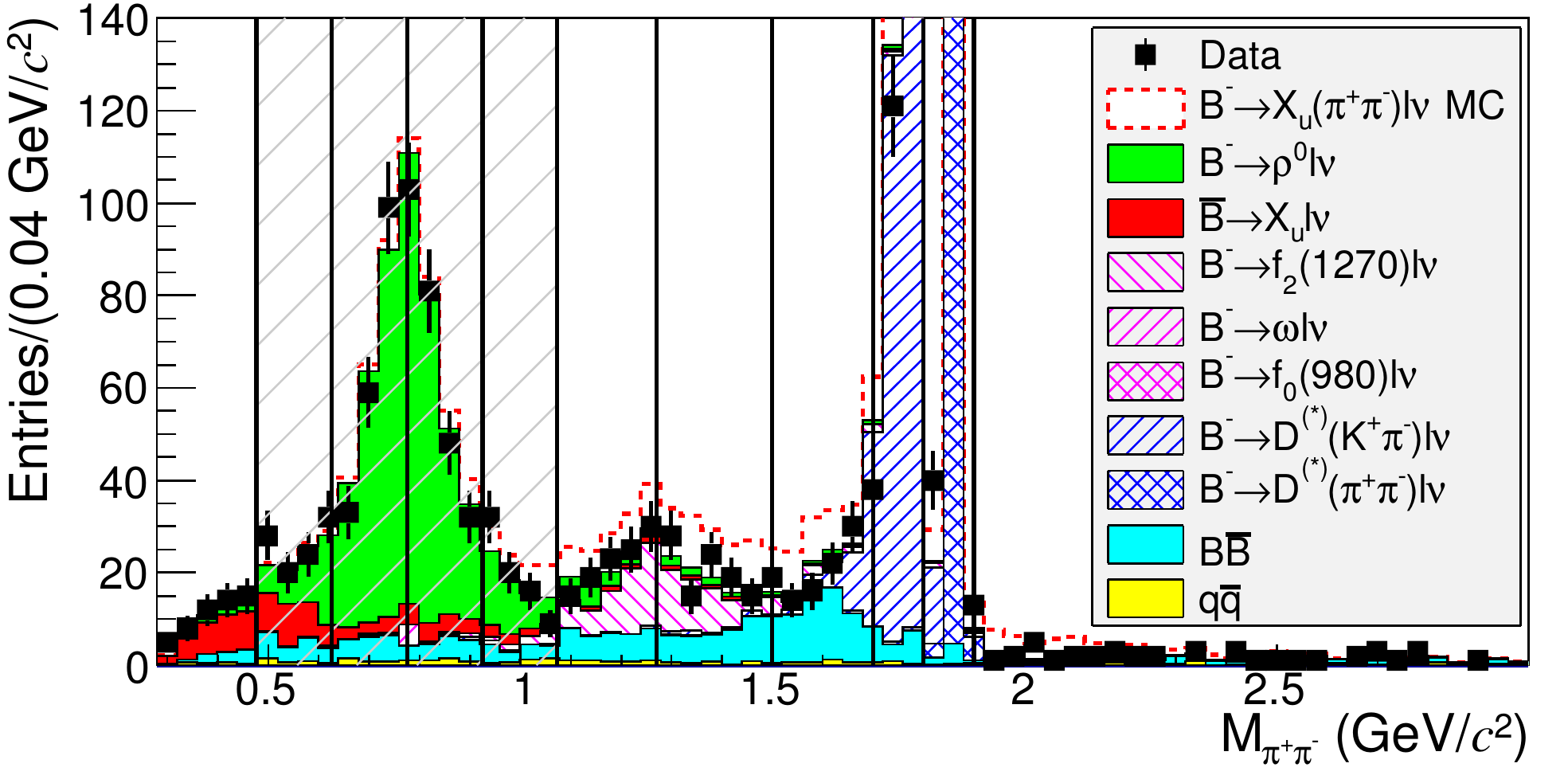}
 \caption{Invariant mass distribution for $\pi^{+}\pi^{-}$ using the hadronic full reconstruction by Belle.}
 \label{fig:massrho}
 \end{figure}
The most recent results for this channel are coming from a Belle tagged analysis~\cite{BelleHad}, whose main results are presented in Table~\ref{tab:rho}. When comparing the branching ratio results with previous measurements (see ref.~\cite{HFAG}), it can be noted that the smallest systematic uncertainty is hold by hadronic tagged measurements, while, contrary to the $\pi$ channel, the untagged measurements have the biggest systematic errors. This behaviour can be explained using the fact that the $\rho$ meson is a wide resonance whose main background is due to other charmless semileptonic $B$ decays. This background is very similar to the signal, for which very harsh kinematic selection are required leading to an increase in the systematic uncertainties. The invariant mass of the $\pi^{+}\pi^{-} $ system is shown if Fig.~\ref{fig:massrho}, for which a peak around $0.8$ GeV corresponds to the $\rho^{0}$ resonance. It can also be noted that the current Monte Carlo scheme overestimates the $\pi^{+}\pi^{-}$ non-resonant contribution, for which more detailed theoretical studies are required. For the first time, Belle has an evidence of a broad resonance around 1.3 GeV dominated by  the $B^{+}\to f_{2}\ell\nu$ decay.

\begin{table}[t]
\begin{center}
\begin{tabular}{l|cccc}  \hline
Theory &  $q^{2}$ [GeV$^{2}$]  &  $N^{\rm fit}$ & $\Delta {\cal B} 10^{-4}$ & $|V_{ub}| 10^{-3}$ \\ \hline
LCSR~\cite{LCSR2}   & $<16$                 & $477\pm 31$    &$1.431\pm 0.091$           &$3.56\pm 0.11\pm 0.09^{+0.54}_{-0.37}$  \\
UKQCD~\cite{UKQCD}  & full range            & $622\pm 35$    &$1.834\pm 0.103$           &$3.68\pm 0.10\pm 0.10^{+0.29}_{-0.34}$  \\  \hline
\end{tabular}
\caption{Results for  $B^{+}\to\rho^{0}\ell^{+}\nu$ reported by the Belle collaboration, the first error is statistical, the second due to systematic uncertainties and the third is the theoretical uncertainty. }
\label{tab:rho}
\end{center}
\end{table}
\subsubsection{$B\to\omega\ell\nu$}
In this section I present two untagged results by BaBaR~\cite{BabarUntaggedW,UntaggedExclusiveBabar} and one tagged measurement by Belle~\cite{BelleHad}. The two untagged measurements by BaBaR differ in the signal selection and background suppression approach. One~\cite{BabarUntaggedW} uses a  neural network discriminator to reduce background and subtract the combinatoric background using a fit the mass sideband data. The other analysis~\cite{UntaggedExclusiveBabar} uses a cut based technique to reduce background and considers the combinatoric background as fit parameter. The tagged measurement by Belle follows the same procedure described earlier for the $B\to\pi\ell\nu$ decay. The results for these measurements are shown in Table~\ref{tab:omega}, where the $\omega $ meson has been reconstructed in the $\omega\to\pi^{+}\pi^{-}\pi^{0}$ decay channel. 
\begin{table}[t]
\begin{center}
\begin{tabular}{l|cccc}  \hline
Experiment &  $q^{2}$ [GeV$^{2}$]  &  $N^{\rm fit}$ & $\Delta {\cal B} 10^{-4}$ & $|V_{ub}| 10^{-3}$ \\ \hline
BaBaR~\cite{BabarUntaggedW}   & full range               & $1125\pm 131$  &$1.21\pm 0.14\pm 0.08$     &$3.23\pm 0.22_{\rm exp}\pm 0.38_{\rm theo}$  \\
        & $0-12$                   & ---            &---                        &$3.37\pm 0.23_{\rm exp}\pm 0.38_{\rm theo}$  \\
BaBaR~\cite{UntaggedExclusiveBabar}   & full range               & $1861\pm 233$  &$1.19\pm 0.16\pm 0.09$     &$3.20\pm 0.10\pm 0.05^{+0.45}_{-0.32}$  \\
Belle~\cite{BelleHad}  & $0-12$                    & $61\pm 11$     &$0.611\pm 0.113$           &$3.08\pm 0.29\pm 0.11^{+0.44}_{-0.31}$  \\  \hline
\end{tabular}
\caption{Results for  $B^{+}\to\omega\ell^{+}\nu$  using the Ball-Zwicky~\cite{LCSR2} calculation for the two untagged BaBaR measurements and the tagged Belle measurement. The first error is statistical, the second due to systematic uncertainties and the third is the theoretical uncertainty. }
\label{tab:omega}
\end{center}
\end{table}

\subsubsection{$B\to\eta\ell\nu$ and $B\to\eta^{\prime}\ell\nu$}
The $\eta$ meson is reconstructed in two decay modes, $\eta\to\gamma\gamma$ and $\eta\to\pi^{+}\pi^{-}\pi^{0}$ and then combine to quote the results. The $\eta^{\prime}$ meson is reconstructed in $\eta^{\prime}\to\eta(\gamma\gamma)\pi^{+}\pi^{-}$. In Table~\ref{tab:eta}, I show the latest results from one untagged measurement form BaBaR and one tagged measurement by Belle.
\begin{table}[t]
\begin{center}
\begin{tabular}{l|cc|cc}  \hline
           &\multicolumn{2}{|c|}{$\eta$}&\multicolumn{2}{|c}{$\eta^{\prime}$}\\
Experiment & $N^{\rm fit}$ & $\Delta {\cal B} \;  10^{-4}$ &$N^{\rm fit}$ & $\Delta {\cal B} \;  10^{-4}$ \\ \hline
BaBaR~\cite{UntaggedExclusiveBabar}      & $867\pm 101$  &$0.38\pm 0.05\pm 0.05$     &$141\pm 49$   &$0.24\pm 0.08\pm 0.03$  \\
Belle      & $39\pm 11$    &$0.42\pm 0.12\pm 0.05$     &$6.1\pm 4.7$  &$<0.57$ at $90\% $ CL  \\  \hline
\end{tabular}
\caption{Results for  $B^{+}\to\eta\ell^{+}\nu$ for one untagged BaBaR measurement and the tagged Belle measurement. The first error is statistical, the second due to systematic uncertainties and the third is the theoretical uncertainty. }
\label{tab:eta}
\end{center}
\end{table}
\section{Search for $B^{-}\to p \bar{p}\ell^{-}\nu$}
\begin{figure}[htb]
 \centering
 \includegraphics[height=2.5in]{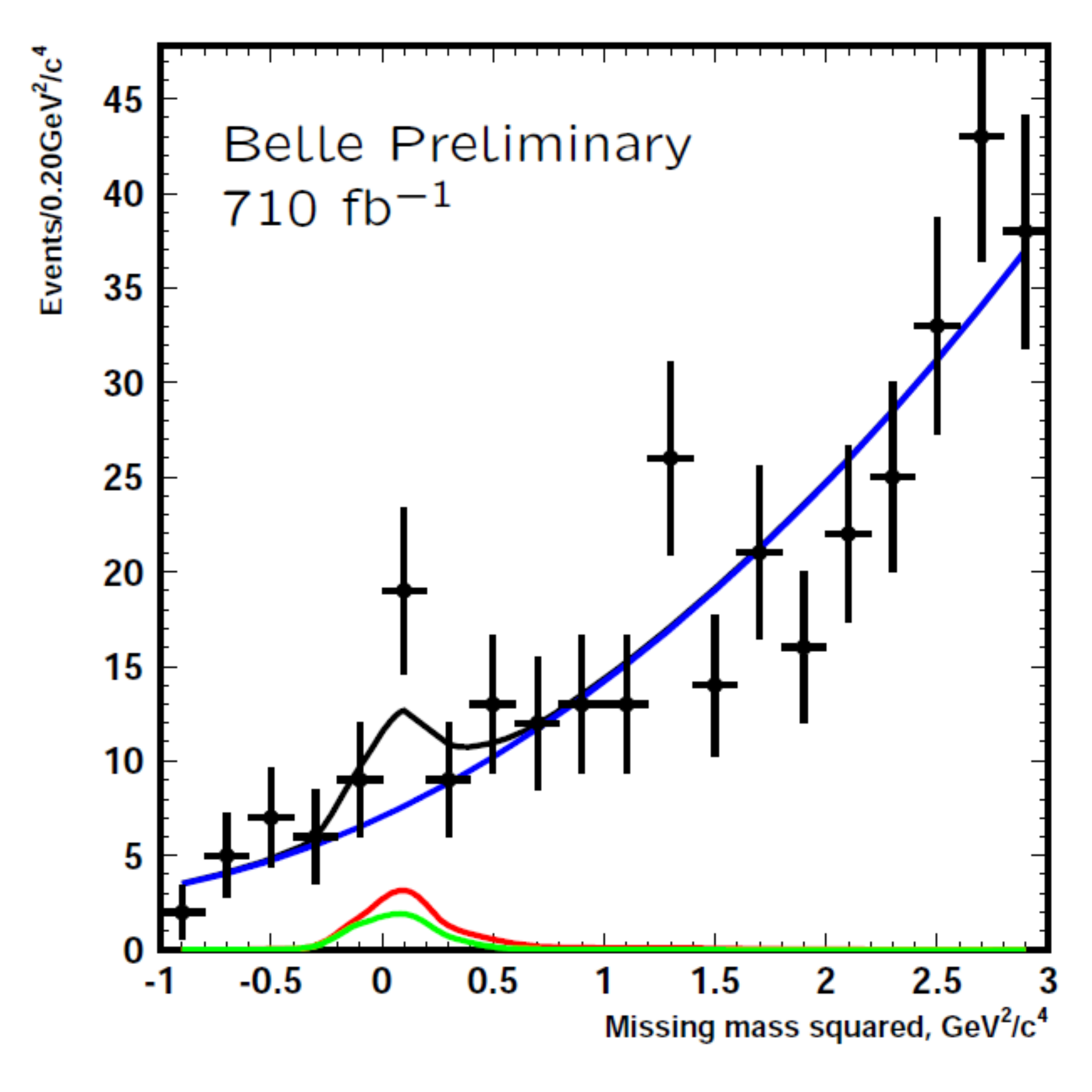}
 \caption{Fitted missing mass squared for the combined channels (electrons and muons) of the $B^{-}\to p \bar{p}\ell^{-}\nu$ decay.}
 \label{fig:ppbar}
 \end{figure}

A phenomenological calculation~\cite{PhenoCalc} suggests that the branching ratio of exclusive semileptonic $B$ decays to a baryon-antibaryon pair is about $10^{-5}-10^{-6}$, making this kind of analysis very challenging with the current data set accumulated by the $B-$factories. However, a recent paper~\cite{PhenoCalc2} estimates this branching ratio to be $(1.04\pm 0.38)\times 10^{-4}$ for $B^{-}\to p \bar{p}\ell^{-}\nu$, which is at the same level of most semileptonic decays with charmless mesons such as $B\to\pi\ell\nu$. This fact is the main motivation for study this decay for the first time in the Belle collaboration. This analysis~\cite{pplnu} uses the complete Belle data set, 711 fb$^{-1}$ of integrated luminosity, with a hadronic full reconstruction of the other $B$ meson. They obtain $18^{+11}_{-9}$ signal events from an ubninned maximum likelihood fit to the $M_{\rm miss}^{2}$, leading to a branching ratio of ${\cal B} (B\to p\bar{p}\ell\nu)=(5.8^{+2.4}_{-2.1}(stat)\pm 0.9(syst)) \times 10^{-6}$ with a significance of $3.2\sigma$. The fitted $M_{\rm miss}^{2}$ distribution is shown in Fig.~\ref{fig:ppbar}. In addition, an upper limit of $9.6\times 10^{-6}$ is estimated at 90$\%$ confidence level. The main source of systematic uncertainties is due to the signal decay model. 

\section{Summary}
New results from Belle and BaBaR using the inclusive and exclusive approach have been presented. Although the results agree between the two collaborations, the tension between the values of $|V_{ub}|$ from inclusive and exclusive measurements still persists. In addition to this, Belle and BaBaR report measurements of $|V_{ub}|$ in other exclusive channels ($B\to\rho,\omega\ell\nu$) different to the traditional $B\to\pi\ell\nu$, which are dominated by theoretical uncertainties. In the near future more detailed analyses of exclusive charmless semileptonic decays with hadron masses above 1 GeV are expected to come out, and thus extend the range of the exclusive measurements and help to reduce systematic uncertainties due to these decays. Finally, new results from the Belle collaboration show evidence of semileptonic decays involving bound states of baryons.     

\end{document}